# Semimetallic carbon allotrope with topological nodal line in mixed $sp^2$-$sp^3$ bonding networks

Ha-Jun Sung[1], Sunghyun Kim[1], In-Ho Lee[2], and K. J. Chang[1*]

[1]Department of Physics, Korea Advanced Institute of Science and Technology, Daejeon 34141, Korea

[2]Korea Research Institute of Standards and Science, Daejeon 34113, Korea

*Corresponding author: kjchang@kaist.ac.kr

**Graphene is known as a two-dimensional Dirac semimetal, in which electron states are described by the Dirac equation of relativistic quantum mechanics. Three-dimensional analogues of graphene are characterized by Dirac points or lines in momentum space, which are protected by symmetry. Here, we report a novel 3D carbon allotrope belonging to a class of topological nodal line semimetals, discovered by using an evolutionary structure search method. The new carbon phase in monoclinic $C2/m$ space group, termed $m$-$C_8$, consists of five-membered rings with $sp^3$ bonding interconnected by $sp^2$-bonded carbon networks. Enthalpy calculations reveal that $m$-$C_8$ is more favorable over recently reported topological semimetallic carbon allotropes, and the dynamical stability of $m$-$C_8$ is verified by phonon spectra and molecular dynamics simulations. Simulated x-ray diffraction spectra propose that $m$-$C_8$ would be one of the unidentified carbon phases observed in detonation shoot. The analysis of electronic properties indicates that $m$-$C_8$ exhibits the nodal line protected by both inversion and time-reversal symmetries in the absence of spin-orbit coupling and the surface band connecting the projected nodal points. Our results may help design new carbon allotropes with exotic electronic**

**properties.**

Carbon, which is one of the most abundant elements in nature, has a rich variety of structural allotropes due to its ability to form *sp*, *sp*$^2$, and *sp*$^3$ hybridized bonds. Graphene, a single layer of carbon in the honeycomb lattice, consists of all-*sp*$^2$ bonds and exhibits the semimetallic band structure with Dirac points. Recently, topological materials including topological insulators (TIs) and topological semimetals (TSMs) have received much attention because of their intriguing physical phenomena and potential applications. The prediction of the TI phase in graphene with spin-orbit coupling (SOC)[1] has stimulated a tremendous amount of theoretical and experimental works to explore new topological materials. The TI state has been evidenced for two-dimensional (2D) HgTe/CdTe quantum wells[2] and three-dimensional (3D) Bi-based chalcogenides[3-5]. The band structure of TIs is characterized by the existence of a bulk band gap as well as gapless boundary states that are protected by the nontrivial topology of bulk electronic states. On the other hand, in TSMs, the valence and conduction bands cross each other at discrete points (Dirac and Weyl semimetals) or along curves (nodal line semimetals) in momentum space. In Dirac semimetals, which have been realized for $Na_3Bi$[6-9] and $Cd_3As_2$[10-14], the band crossing points at the Fermi energy have fourfold degeneracy. By breaking either inversion or time-reversal symmetry in Dirac semimetals, one can obtain Weyl semimetals in which each Dirac point splits into a pair of doubly degenerate Weyl points with opposite chirality. Weyl semimetals have been proposed for compounds containing heavy elements, such as pyrochlore iridates[15], $HgCr_2Se_4$[16], and transition metal dichalcogenides[17,18], and the prediction of the Weyl semimetal state in the TaAs family[19,20] has been verified by several experiments[21-24].

In nodal line semimetals, the formation of Dirac nodes along a closed loop or line requires inversion and time-reversal symmetries without SOC in $Cu_3$(Pd, Zn)N[25,26], $Ca_3P_2$[27], and alkaline-earth metals (Ca, Sr, Yb)[28] and compounds $AX_2$ (A = Ca, Sr, Ba; $X$ = Si, Ge, Sn)[29]. When SOC is included, additional non-symmorphic symmetry is necessary to protect the nodal line against gap opening in ZrSi(S, Se, Te)[30-32] and IrF$_4$[33]. In non-centrosymmetric PbTaSe$_2$, the nodal line is protected by mirror reflection symmetry even in the presence of SOC[34,35]. Meanwhile, the TSM phase was also reported for 3D carbon networks constructed from graphene, such as Mackay-Terrenes crystal[36], interpenetrated graphene network (IGN)[37,38], and bco-C$_{16}$[39]. Due to a negligible SOC[40], the nodal lines of these semimetallic carbon allotropes are protected by a combination of inversion and time-reversal symmetries. Recent experiments have demonstrated the synthesis of 3D graphene networks with high electrical conductivity by a chemical vapor deposition technique, while its crystal structure is not identified yet[41]. Other 3D metallic carbon allotropes were proposed, including $H$18 carbon[42] and $T$6 carbon[43] in mixed $sp^2$-$sp^3$ bonding networks and ThSi$_2$-type tetragonal bct4 carbon[44] and $H$6 carbon[45] with all-$sp^2$ bonding. However, the band overlap occurs at different crystal momenta in $H$18 and $T$6 carbon[42,43], whereas the metallic nature of bct4 and $H$6 carbon arises from the twisted $\pi$ states that make these allotropes dynamically unstable[46].

In this work, we report a new carbon allotrope belonging to a class of topological nodal line semimetals, based on global optimization and first-principles density functional calculations. The new carbon phase, termed $m$-C$_8$, consists of five-membered rings with $sp^3$ hybridized bonds and $sp^2$-bonded carbon networks, and the enthalpy of $m$-C$_8$ is lower than those of recently proposed carbon allotropes with topological nodal lines. The dynamical stability of $m$-C$_8$ is verified by phonon spectra and molecular dynamics simulations. Based on the analysis of x-ray diffraction spectra and enthalpy-pressure curves, we propose that $m$-C$_8$ can be present in detonation soot[47] and a phase transition from graphite to $m$-C$_8$ can occur under pressure.

**Results and Discussion**

**Structure and stability of a new carbon allotrope.** We explored metallic carbon allotropes with $sp^2$-$sp^3$ hybridized bonds by using a computational search method (Methods). Among many low-energy allotropes, we obtained a very distinctive crystal structure in the $C2/m$ space group (Fig. 1a), especially for the $N = 8$ system. For the $N = 16$ system, the same low-energy allotrope was also found by the computational search method, while both the number of atoms per unit cell and the cell volume are doubled. The monoclinic $C2/m$ structure, denoted as $m$-$C_8$, has the equilibrium lattice parameters, $a = 7.010$ Å, $b = 2.480$ Å, $c = 6.608$ Å, and $\beta = 71.2°$, and four inequivalent Wyckoff positions, $4i$ (0.272, 0, 0.446), (0.632, 0, 0.79), (0.781, 0, −0.072), and (0.007, 0, 0.878), occupied by the $C_1$, $C_2$, $C_3$, and $C_4$ atoms, respectively. The $m$-$C_8$ allotrope is characterized by five-membered carbon rings interconnected by graphitic carbon networks. The graphitic networks are composed of $sp^2$-bonded $C_1$ atoms, with the bond length of $d_{C_1-C_1} = 1.412$ Å, whereas the $C_2$, $C_3$, and $C_4$ atoms forming five-membered rings are all $sp^3$-bonded, with the bond lengths of $d_{C_2-C_3} = 1.553$ Å, $d_{C_2-C_4} = 1.527$ Å, $d_{C_3-C_3} = 1.540$ Å, $d_{C_3-C_4} = 1.514$ Å, and $d_{C_4-C_4} = 1.570$ Å. Note that the bond length between the $C_1$ and $C_2$ atoms, which connect graphitic sheets to five-membered rings, is $d_{C_1-C_2} = 1.493$ Å, lying in between those of graphite (1.420 Å) and diamond (1.544 Å).

In Table 1, the calculated equilibrium volume, lattice parameters, bond lengths, and total energy of $m$-$C_8$ are compared with those for diamond, graphite, and several recently reported metallic allotropes, $T6$ carbon[43] and IGN[37] in mixed $sp^2$-$sp^3$ bonding networks and oC8 carbon[48] and bco-$C_{16}$[39] in all-$sp^2$ bonding networks. The equilibrium volume of $m$-$C_8$ is 6.80 Å$^3$ per atom, placing in between those of graphite and diamond. Because of the mixture of $sp^2$ and $sp^3$ hybridized bonds, the $m$-$C_8$ allotrope has four different bond angles of 93.7°, 107.3°, 114.1°, and 121.3°, which deviate from the ideal bond angles of graphite (120°) and diamond (109.5°). Owing to the induced strain, the $m$-$C_8$ structure has the excess energy of 0.22 eV/atom compared with the diamond phase. On the other hand, $m$-$C_8$ is more stable by 0.13–0.29 eV/atom than $T6$ carbon, oC8 carbon, and bco-$C_{16}$. It is interesting to note that, although the total energy of $m$-$C_8$ at the equilibrium volume is higher by 0.05 eV/atom than that of IGN, its enthalpy is lower for pressures above 10 GPa, as shown in Fig. 2. From the enthalpy vs pressure curve, a possible synthesis of $m$-$C_8$ is expected under compression of graphite. It was suggested that graphite may transform to oC8 carbon, which is a denser form of bco-$C_{16}$, above 65 GPa[39,48]. However, our calculations indicate that $m$-$C_8$ is lower in enthalpy than bco-$C_{16}$ up to 77 GPa, and a transition from graphite to $m$-$C_8$ is more likely to occur at a lower pressure of 60 GPa.

We examined the stability of $m$-$C_8$ by calculating the full phonon spectra and found no imaginary phonon modes over the entire Brillouin zone (BZ) (Fig. 1b), indicating that $m$-$C_8$ is dynamically stable. In addition, we carried out first-principles molecular dynamics (MD) simulations at a high temperature of 1500 K. For a $3 \times 2 \times 2$ supercell containing 96 atoms, we confirmed that the $m$-$C_8$ allotrope is stable up to 100 ps (Fig. 1c). Owing to the thermal stability, the synthesis of $m$-$C_8$ is highly expected under high pressure as well as high temperature. We also calculated the elastic constants of $m$-$C_8$, and confirmed that the elastic constants meet the criteria for mechanical stability in monoclinic structure[49].

The x-ray diffraction (XRD) spectra of $m$-$C_8$ were simulated and compared with the experimental data from detonation soot (sample Alaska B)[47], along with those of graphite, diamond, $T6$ carbon, bco-$C_{16}$, and IGN in Fig. 3. In the detonation soot, the prominent peaks around 26.5° and 43.9° are attributed to the graphite (002) and diamond (111) diffractions, respectively. The (101) peak of $T6$ carbon, (101) peak of bco-$C_{16}$, and (001) peak of IGN match with the experimental XRD data located at 37.4°, 30.0°, and 21.4°, respectively. However, the low angle peak at 13.4° does not match any previously reported carbon phases. This peak was also observed in different detonation soot[47], providing that an unknown carbon phase should be produced. Our simulated XRD results show that the main (001) peak of $m$-$C_8$ well explains the unidentified peak at 13.4°. Moreover, the (111), (002), and (112) peaks of $m$-$C_8$ match the experimental XRD spectra located at 25.1°, 27.8°, and 32.1°, respectively, indicating the presence of $m$-$C_8$ in the specimen produced by detonation experiments[47].

**Band structure of a new carbon allotrope.** Finally, we examined the band topology of $m$-$C_8$. In Fig. 4a, the band structure exhibits linear dispersions around the Fermi level where the valence and conduction bands touch each other, similar to graphene. The linear bands are mainly derived from the $sp^2$ hybridized C atoms in graphitic sheets, as illustrated in the distribution of charge densities (see Fig. 4b), while more dispersive bands far from the Fermi level are associated with the $sp^3$ hybridized bonds in five-membered rings. From the band structure in the full BZ, we find that the crossing points of the valence and conduction bands form a continuous nodal line piercing the extended BZ, without any interference with $sp^3$-hybridized bands (Figs. 4 and 5). Thus, the $m$-$C_8$ allotrope with the symmorphic symmetry belongs to a class of topological nodal line semimetals[25-39]. Recently, two types of topological nodal line semimetals were proposed, depending on systems with and without SOC[50,51]. In the former, since the SOC may open up gaps at the band crossing points, both inversion and time-reversal symmetries are insufficient to protect the band crossings, while an additional non-symmorphic symmetry can protect the nodal line[52,53]. In carbon systems like $m$-$C_8$, with a negligible SOC, the topological nodal line survives by a combination of inversion and time-reversal symmetries.

Based on the analysis of parity eigenvalues, Fu and Kane proposed the $Z_2$ topological invariants $(v_0;v_1v_2v_3)$ to describe the topological nature of TIs[52,53]. Similarly, the parities of energy states can be used to assign the $Z_2$ topological invariants in topological nodal line semimetals. For the eight time-reversal invariant momenta (Fig. 5a), the products of the parity eigenvalues ($\delta$) for the occupied bands are listed in Table 2. We find that $m$-$C_8$ is characterized by the weak $Z_2$ indices (0;111) due to the value of $\delta = -1$ at the Y and Z points. Since $m$-$C_8$ has no mirror reflection symmetry, the bulk nodal line does not appear in a mirror-invariant plane. To visualize the formation of topological surface states, we calculated the surface band structure for a slab geometry composed of 20 graphitic layers, where the (110) surface is exposed to vacuum. As the bulk BZ is projected onto the (110) surface BZ, one can expect that the nodal line is located near the $\overline{X}$ and $\overline{S}$ points (Fig. 5a). In fact, the projected band structure clearly shows the formation of the nearly flat surface state connecting the projected nodal points around the $\overline{X}$ point (Fig. 5b).

In summary, we have predicted a novel carbon allotrope $m$-$C_8$ in mixed $sp^2$-$sp^3$ bonding networks by using the evolutionary structure search method. The monoclinic structure of $m$-$C_8$ is composed of five-membered rings connected by graphitic sheets. The stability of $m$-$C_8$ is verified by calculating the full phonon spectra and MD simulations at the high temperature of 1500 K. From the analysis of the electronic band structure, we have identified that $m$-$C_8$ belongs to the class of topological nodal line semimetals, exhibiting the topological nodal line in bulk and the topological surface states at surface boundaries. Since the SOC is extremely weak in $m$-$C_8$, the nodal line is protected by the coexistence of inversion and time-reversal symmetries. While it remains a challenge to synthesize the crystalline form of $m$-$C_8$, our results provide not only a perspective of the novel electronic structure of carbon allotropes but promote future studies to explore new carbon allotropes with exotic electronic and transport properties.

**Methods**

**Computational structure searches and electronic structure calculations**

We explored new carbon allotropes with $sp^2$-$sp^3$ hybridized bonds by using a computational search method[54], in which the conformational space annealing (CSA) algorithm[55] for global optimization is combined with first-principles electronic structure calculations. The efficiency of this approach has been demonstrated by successful applications to predict Si and C allotropes with direct band gaps[56-58]. For various carbon systems with $N$ atoms per unit cell ($N$ = 8, 12, 16, 20), we optimized the degrees of freedom, such as atomic positions and lattice parameters, with the number of configurations setting to 40 in the population size of CSA.

The objective function was designed to prevent carbon structures from forming either all-$sp^2$ or all-$sp^3$ hybridized bonds, such as graphite and diamond, so that mixed $sp^2$-$sp^3$ bonding networks were promoted. The enthalpy minimization and electronic structure calculations were performed within the framework of density functional theory. We used the functional form proposed by Armiento and Mattsson (AM05)[59] for the exchange-correlation potential and the projector augmented wave potentials[60], as implemented in the VASP code[61]. In graphite, interlayer interactions are described more accurately with the AM05 functional, compared with other functional forms for the exchange-correlation potential. The wave functions were expanded in plane waves up to an energy cutoff of 600 eV. At the final stage of optimization, we used an even higher energy cutoff of 800 eV.

**References**

1. Kane, C. L. & Mele E. J. Quantum spin Hall effect in graphene. *Phys. Rev. Lett.* **95**, 226801 (2005).

**Acknowledgments** This work was supported by Samsung Science and Technology Foundation under Grant No. SSTF-BA1401-08.

.

**Table 1 | Calculated structural and electronic properties of various carbon allotropes**

| | Method | $V_0$ | $a$ | $b$ | $c$ | $d_{C-C}$ | $E_{tot}$ | $E_g$ |
|---|---|---|---|---|---|---|---|---|
| Diamond | AM05 | 5.60 | 3.554 | | | 1.538 | -9.467 | 5.36 |
| | Exp.[62] | 5.67 | 3.567 | | | 1.544 | | 5.47 |
| Graphite | AM05 | 8.79 | 2.461 | | 6.706 | 1.421 | -9.494 | 0 |
| | Exp.[63] | 8.78 | 2.460 | | 6.704 | 1.420 | | 0 |
| $T6$ | AM05 | 6.76 | 2.600 | | 6.000 | 1.340, 1.540 | -8.962 | 0 |
| IGN | AM05 | 7.53 | 2.456 | | 4.291 | 1.403, 1.519 | -9.302 | 0 |
| bco-$C_{16}$ | AM05 | 7.71 | 7.806 | 4.875 | 3.239 | 1.383~1.458 | -9.120 | 0 |
| oC8 | AM05 | 6.48 | 7.788 | 2.499 | 2.665 | 1.347~1.632 | -9.115 | 0 |
| $m$-$C_8$ | AM05 | 6.80 | 7.010 | 2.480 | 6.608 | 1.412~1.538 | -9.251 | 0 |

Calculated equilibrium volumes ($V_0$ in Å$^3$/atom), lattice parameters ($a$, $b$, and $c$ in Å), bond lengths ($d_{C-C}$ in Å), total energies ($E_{tot}$ in eV/atom), and band gaps ($E_g$ in eV) for diamond, graphite, $T6$ carbon, IGN, bco-$C_{16}$, oC8, and $m$-$C_8$.

**Table 2 | Parity eigenvalues of $m$-$C_8$**

| TRIM | $\delta$ | TRIM | $\delta$ |
|---|---|---|---|
| $\Gamma(\Lambda_{0,0,0})$ | +1 | $N(\Lambda_{1/2,0,0})$ | +1 |
| $Y(\Lambda_{1/2,1/2,0})$ | −1 | $N'(\Lambda_{0,1/2,0})$ | +1 |
| $Z(\Lambda_{0,0,1/2})$ | −1 | $M(\Lambda_{1/2,0,1/2})$ | +1 |
| $L(\Lambda_{1/2,1/2,1/2})$ | +1 | $M'(\Lambda_{0,1/2,1/2})$ | +1 |

Product of parity eigenvalues ($\delta$) for the occupied bands at the time-reversal invariant momenta (TRIM) in $m$-$C_8$.

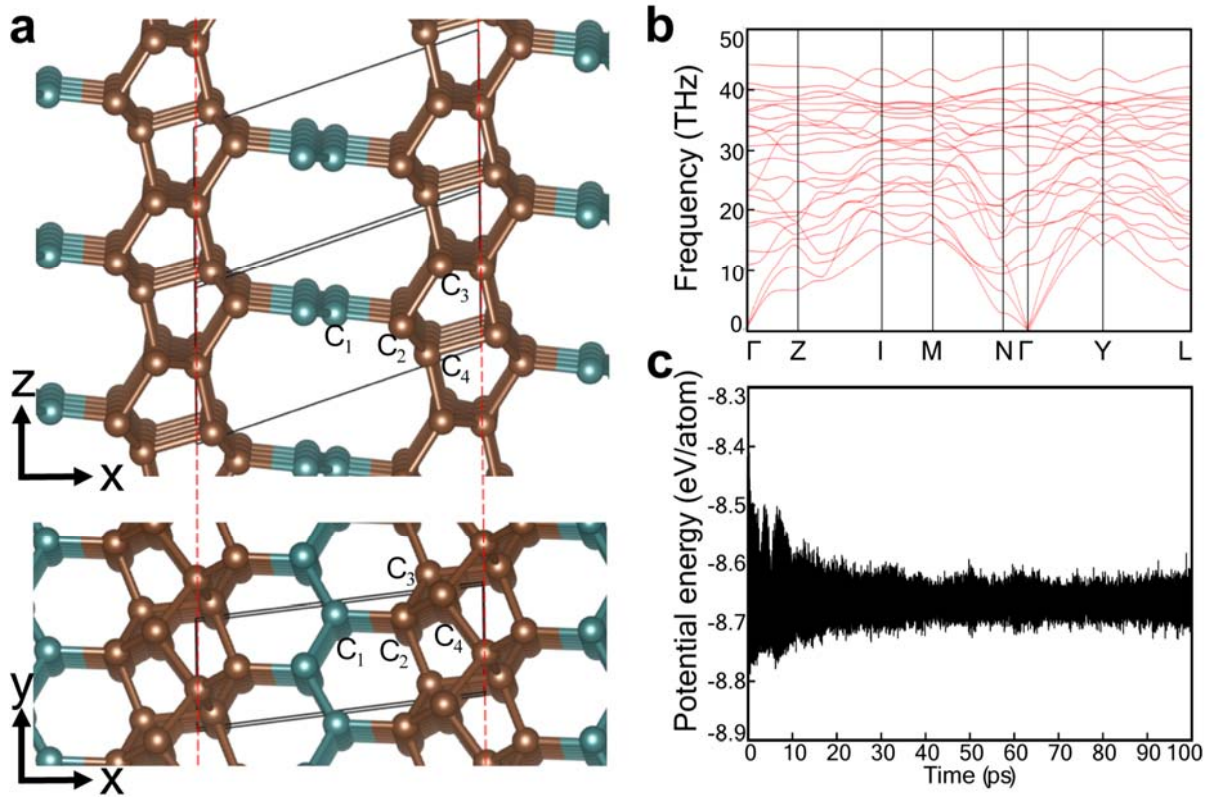

**Figure 1 | Atomic structure of *m*-$C_8$ and its dynamical and thermal stability.** (a) Side and top views of the atomic structure of *m*-$C_8$ in *C*2/*m* space group. The lattice parameters in monoclinic structure are *a* = 7.010 Å, *b* = 2.480 Å, *c* = 6.608 Å, *β* = 71.2°, and $C_1$, $C_2$, $C_3$, and $C_4$ denote four inequivalent Wyckoff positions. (b) Calculated phonon spectra of *m*-$C_8$ at zero pressure and (c) potential energy fluctuations during MD simulations at 1500 K.

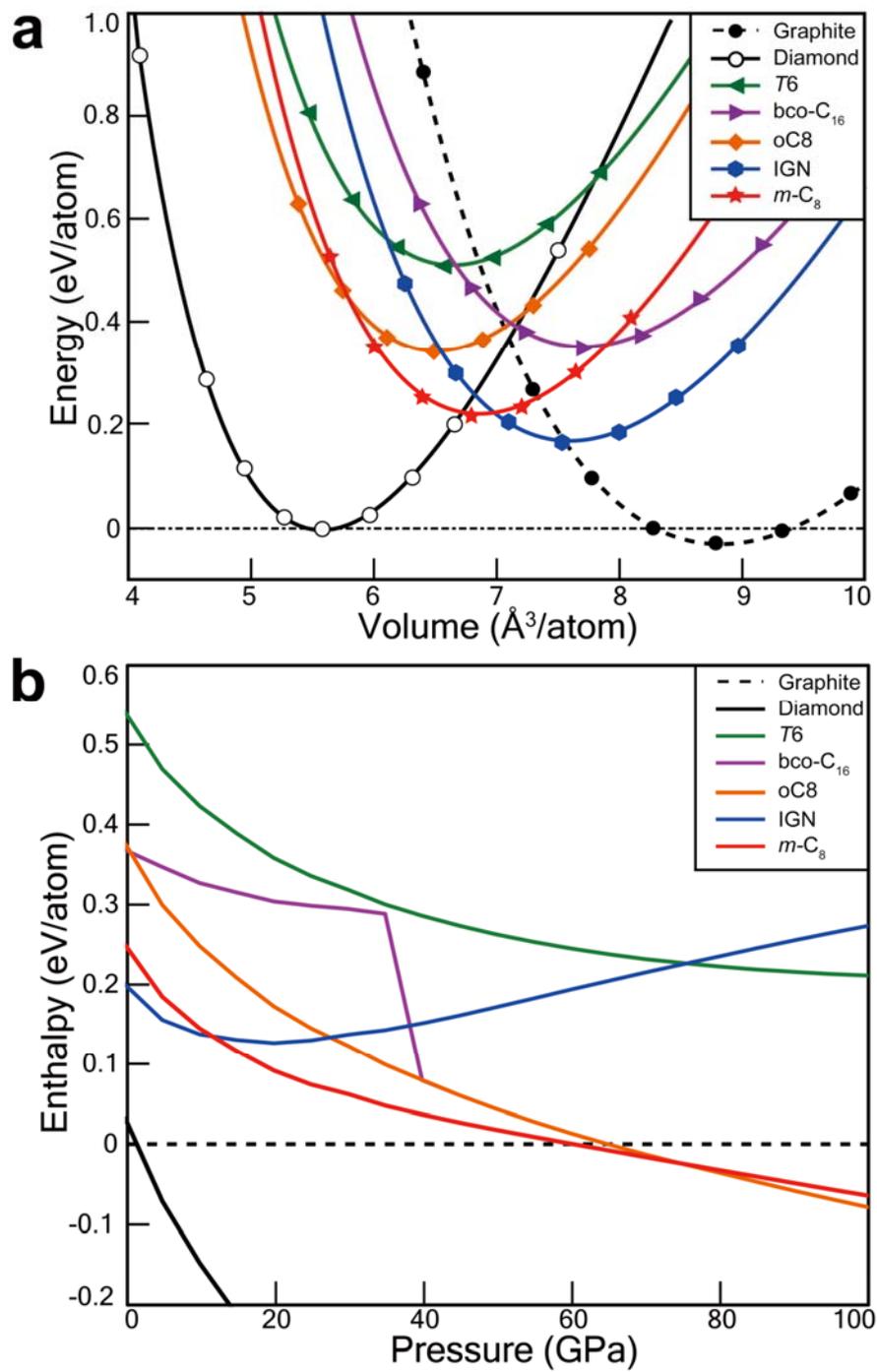

**Figure 2 | Energy vs volume and enthalpy vs pressure curves.** (a) Total energy as a function of volume and (b) enthalpy as a function of pressure curves for diamond, graphite, *T*6 carbon, IGN, bco-$C_{16}$, oC8, and *m*-$C_8$. The transition from graphite to *m*-$C_8$ occurs at a pressure of 60 GPa.

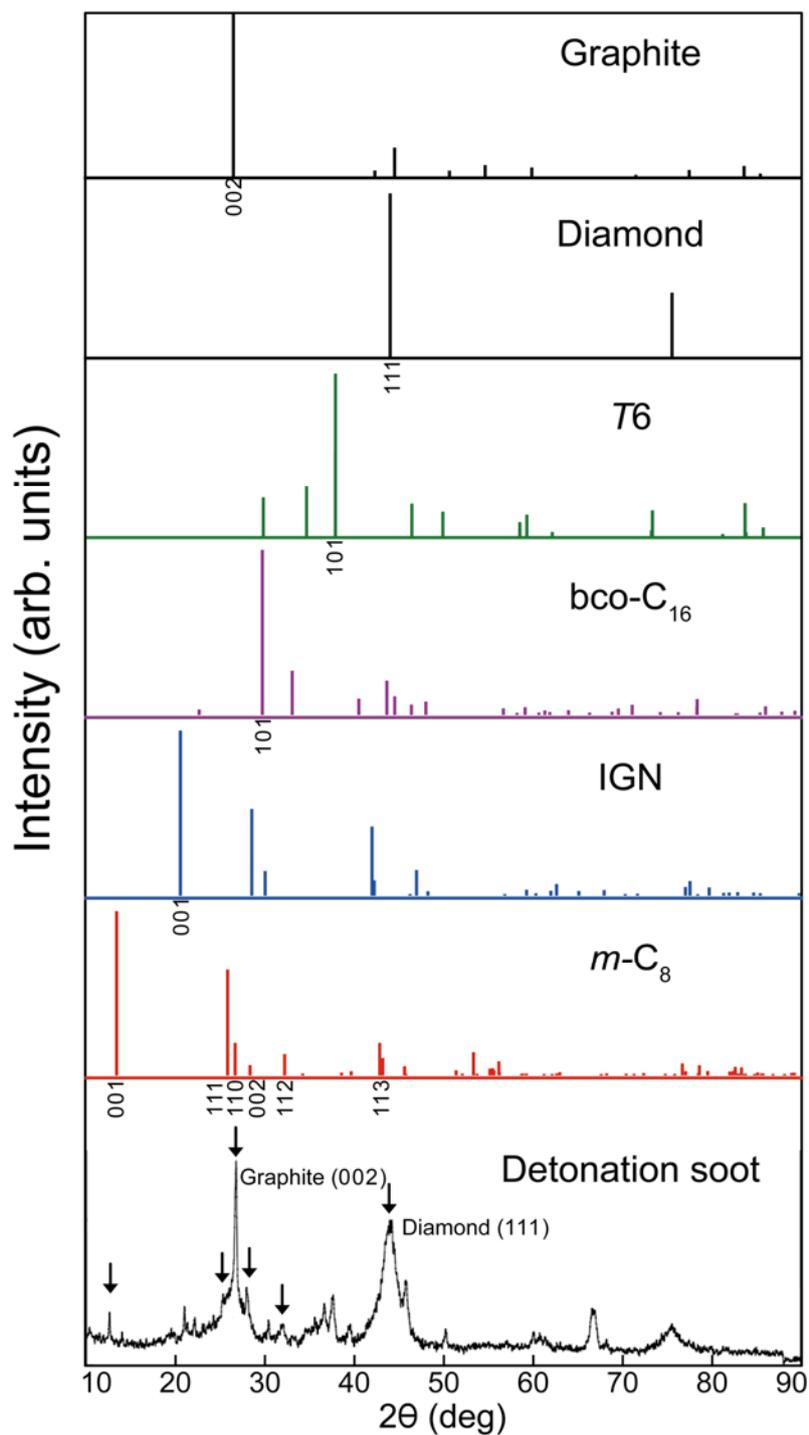

**Figure 3 | Simulated x-ray diffraction (XRD) patterns.** The simulated XRD spectrum for graphite, diamond, $T6$ carbon, bco-$C_{16}$, IGN, and $m$-$C_8$ are compared with those experimentally observed for detonation soot of TNT (sample Alaska B)[47]. Arrows indicate the XRD peaks related to $m$-$C_8$.

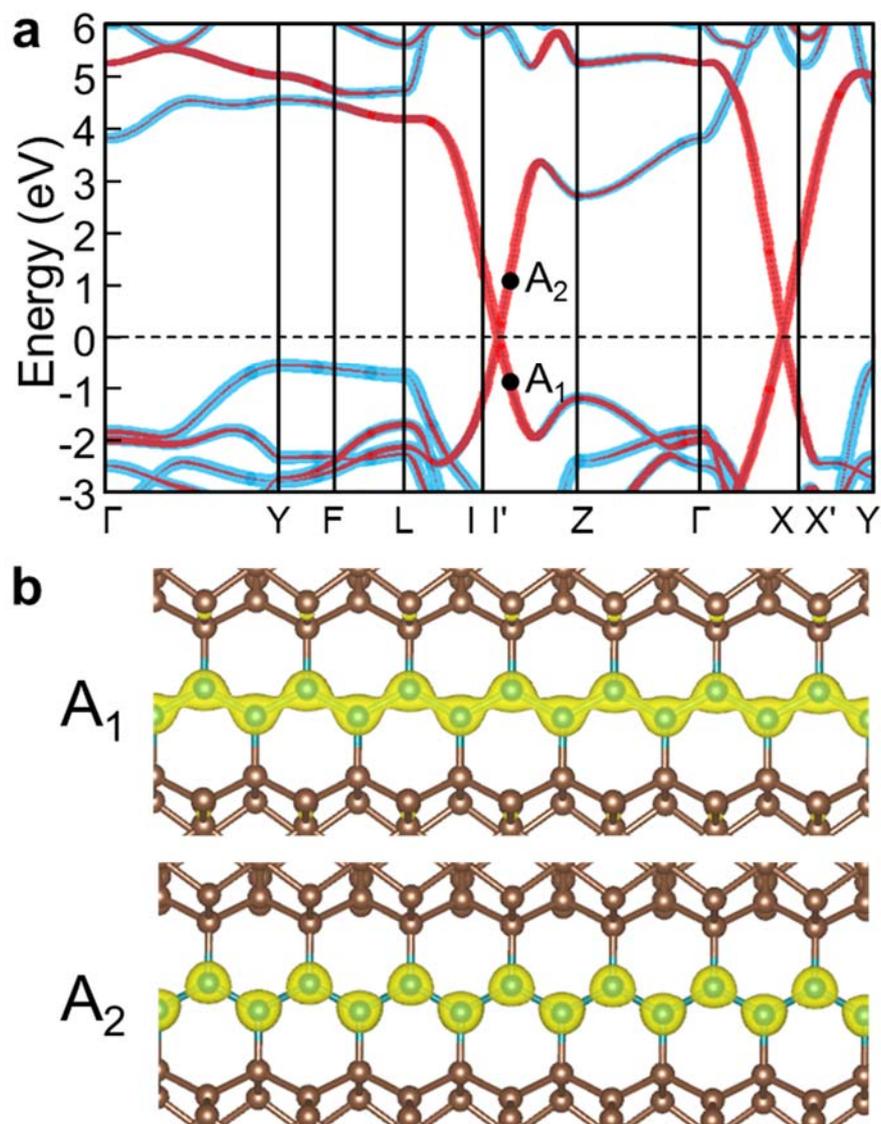

**Figure 4 | Electronic structure of *m*-C$_8$.** (**a**) Band structure of *m*-C$_8$. The thickness of red and blue colored bands represents the degree of charge density distributions in the *sp$^2$* and *sp$^3$* hybridized atoms, respectively. (**b**) Distribution of the charge densities for the linear bands (A$_1$ and A$_2$) near the Fermi energy which are mainly derived from the carbon chains in *sp$^2$* bonding networks.

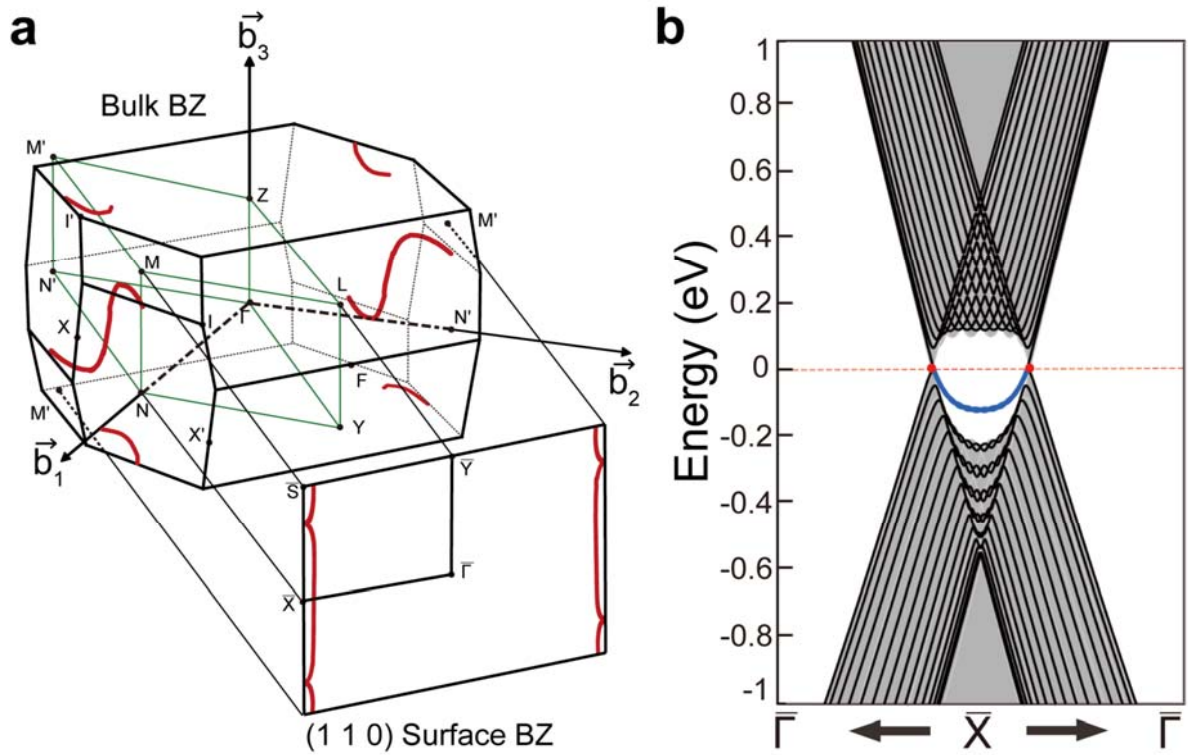

**Figure 5 | Nodal lines and surface states.** (**a**) The 3D Brillouin zone (black polyhedron) with several high-symmetry momenta and the nodal lines (red lines) at the Fermi energy in the monoclinic structure of *m*-C$_8$. TRIM points (vertices in green rhombohedral) and their projection onto the (110) surface BZ are indicated. (**b**) Topologically protected (110)-surface band (blue line) nestled inside the bulk nodal line (red dots).